\documentclass[aps,prd,nofootinbib]{revtex4}
\usepackage{amsmath}
\usepackage{graphicx}
\usepackage{dcolumn}
\usepackage{bm}
\usepackage{amssymb}
\usepackage{latexsym}

\def\be{\begin{equation}}
\def\ee{\end{equation}}
\def\bea{\begin{eqnarray}}
\def\eea{\end{eqnarray}}

\begin{document}

\title{Constraints on Dark Energy Models Including Gamma Ray Bursts}

\author{Hong Li$^{1}$, Meng Su$^{1}$, Zuhui
Fan$^{1}$, Zigao Dai$^{2}$ and Xinmin Zhang$^{3}$}

\affiliation{${}^1$ Department of Astronomy, School of Physics,
Peking University, Beijing, 100871, P. R. China}

\affiliation{ ${}^2$ Department of Astronomy, Nanjing University,
Nanjing 210093, P. R. China}

\affiliation{${}^3$Institute of High Energy Physics, Chinese
Academy of Science, P.O. Box 918-4, Beijing 100049, P. R. China}

\begin{abstract}
In this paper we analyze the constraints on the property of dark
energy from cosmological observations. Together with SNe Ia Gold
sample, WMAP, SDSS and 2dFGRS data, we include 69 long Gamma-Ray
Bursts (GRBs) data in our study and perform global fitting using
Markov Chain Monte Carlo (MCMC) technique. Dark energy perturbations
are explicitly considered. We pay particular attention to the time
evolution of the equation of state of dark energy parameterized as
$w_{DE}=w_0+w_a(1-a)$ with $a$ the scale factor of the universe,
emphasizing the complementarity of high redshift GRBs to other
cosmological probes. It is found that the constraints on dark energy
become stringent by taking into account high redshift GRBs,
especially for $w_a$, which delineates the evolution of dark energy.
\end{abstract}

\maketitle

\section{Introduction}
Cosmological observations, including Type Ia supernovae (SNe
Ia)\cite{1997snia,1998snia,Riess,Riess:2006}, Cosmic Microwave
Background radiation (CMB)\cite{Spergel,Spergel:2006hy}, Large-Scale
Structures (LSS)\cite{Seljak} and so on, have provided strong
evidence for a spatially flat universe being in a stage of
accelerating expansion. In the context of Friedmann-Robertson-Walker
cosmology, this acceleration is attributed to a new form of energy
with negative pressure, dubbed dark energy (DE), whose nature
remains a big puzzle. The simplest candidate for DE is the
cosmological constant.  However, it suffers from the notorious fine
tuning and coincidence problems \cite{SW89,ZWS99}. Many dynamical
models on DE, such as quintessence
\cite{quint1,quint2,quint3,quint4}, phantom \cite{phantom},
k-essence \cite{kessence1,kessence2} and quintom, have been proposed
to avoid above difficulties. Among them, the quintom model allows
the equation of state (EOS) crossing -1 during evolution
\cite{quintom}. Extensive studies on this model has been carried on
both theoretically and phenomenologically
\cite{quintom,Feng:2004ff,xia,Xia:2005ge,Xia:2006cr,Zhao:2005vj,
Zhao:2006bt,Li:2005fm,Zhang:2005eg,Zhang:2006ck,
Guo:2004fq,weihao05,cai0609039,Andrianov05,
zhangxin,McInnes05,Aref'eva05,Grande06,
Chimento06,Cannata06,Lazkoz06,Stefancic05}.

Different DE models predict different global evolutions as well as
different structure formations of the universe. Therefore
cosmological observations can provide important constraints on the
nature of DE. However, degeneracies of cosmological parameters exist
in almost all cosmological observations, i.e., they are sensitive
not to single parameters but to some specific combinations of them.
It is therefore highly necessary to combine different probes to
break parameter degeneracies so as to achieve tight cosmological
constraints. Furthermore, different observations are affected by
different systematic errors, and it is thus helpful to reduce
potential biases by combining different probes. To perform joint
analyzes, global fitting is the most secure study, because it avoids
using strong, and sometimes even inappropriate priors obtained from
other observations.

Observations on SNe Ia have played important roles in DE
studies\cite{1998snia,Riess,Riess:2006}. However, due to the
difficulties of detecting high-redshift SNe Ia, such a geometrical
measurement is limited. The future SNAP/JDEM is planed to probe up
to redshift around $z\sim 2$. On the other hand, the 3-year WMAP
data (WMAP3), which is so far the most precise measurement of the
CMB, reveals the information obout our universe at redshift around
$z\sim 1100$. In between, there have not been many probes accessible
to us. In this regard, GRBs are the most promising tracers of the
evolution of the universe at redshift around a few to even a few
tens because they are the most powerful events in the universe. The
currently operating Swift satellite is able to detect about $100$
long duration GRB events within one year, and we expect that it will
open up a potentially new window for GRB cosmographic studies. At
present, the measured redshifts of GRBs have extended to
$z=6.29$\cite{kawai}. Even though the physical origin of the long
GRB is not very clear, many extensive discussions about the
relations between the spectral and temporal parameters show the
potential for using long GRBs as cosmic candles for cosmography.

Recently, headway has been made in considering how to make GRBs
standard candles. In the literature there are many studies about the
intrinsic correlations between temporal or spectral properties of
GRBs and their isotropic energies and luminositiy, for example, the
spectral lag $-$ the luminosity correlation
($\tau-L_{iso}$)\cite{tau-liso}, the luminosity $-$ variablity
correlation (V-$L_{iso}$)\cite{v-liso}, the spectral peak energy $-$
isotropic energy correlation ($E_{peak}$-$E_{iso}$)\cite{ep-eiso},
the peak spectral energy$-$isotropic luminosity correlation
($E_{peak}$-$L_{iso}$)\cite{ep-liso} and the isotropic luminosity
$-$ peak energy $-$ high signal time scale correlation
($L_{iso}$-$E_{peak}$-$T_{0.45}$)\cite{liso-ep-t} and so on. These
correlations help GRBs nearly to be "known candles"
\cite{firmani0610248}. Shaefer obtained the first GRB Hubble diagram
of 9 GRBs with known redshift by using the spectral lag and the
variability indicators, and from the GRB hubble diagram he
constrained $\Omega_{m}< 0.35$ within $1 \sigma$ confidence
level\cite{1st-hubble diagram}. After that, many investigations have
been triggered on using GRBs as cosmological probes
\cite{dailiangxu2004,ghirlanda2004a,Firmani2005,friedman2005,
lamb2005,liang enwei2005,mortsell2005,xudong2005,sumeng0611155}. 
Very recently, Shaefer\cite{shafer2006} has constructed a GRB Hubble
Diagram with 69 GRBs over a redshift range of $0.17$ to $> 6$, with
39 GRBs having redshifts $z>1.5$. It is the largest GRB sample so
far. The aim of this paper is to present a more general analysis on
the EOS of DE by including GRB data in addition to CMB, LSS and SNe
Ia in global fitting. We mainly consider the dynamical dark energy
parameterization for there is no compelling reason to neglect the
evolution of dark energy factitiously. We employ MCMC techniques for
the analysis. The MCMC method is the global fitting on the
cosmological parameters and we use the original CMB and LSS data
rather than use the CMB shift parameter, the linear growth factor or
the Baryon Acoustic Oscillation (BAO) measured from the Large scale
structure survey. The global fitting is the most secure way for
processing data because it is a joint analysis and it can avoid
using some strong or even inappropriate priors from others. The
effects of dark energy perturbations are carefully taken into
account with great attention paid to the perturbations when the
equation of state gets across -1. Our paper is structured as
follows: in Section II we describe the method and the data; in
Section III we present our results on the determination of
cosmological parameters from global fitting; finally we present our
conclusions in Section V.

\section{Method and data}

In this section we describe the method used in the fitting process.
For the DE parametrization, we adopt $\Lambda CDM$ model, constant
$w$ and pay particular attention to the commonly used EOS of the
form\cite{Linderpara}:
\begin{equation}
\label{EOS} w_{DE}(a) = w_{0} + w_{a}(1-a)
\end{equation}
where $a=1/(1+z)$ is the scale factor and $w_{a}$ characterizes the
``running" of the EOS.
The evolution of DE density is then
\begin{equation}
\rho_X(a)/\rho_X(a_0=1)=a^{-3(1+w_0+w_a)} \exp{[3w_a(a-1)]}~.
\end{equation}

For DE models whose EOS is not equal to $-1$, the perturbation
inevitably exists. The perturbation of DE has no effect on the
geometric constraints of SN Ia, however, for the CMB and LSS data,
the perturbation of DE should be considered, because the late time
ISW effects differ significantly when DE perturbation are
considered, and the ISW effects take an important part on large
angular scales of CMB and the matter power spectrum of
LSS\cite{Zhao:2005vj}. For quintessence-like or phantom-like models
where $w$ does not cross $-1$, there are not fundamental
difficulties in dealing with perturbations. In parameter fitting,
however, biases may be introduced if we limit our considerations
only to quintessence or only to phantom models. Thus it is more
natural and consistent to allow $w$ crossing $-1$ in the fitting
analysis. Furthermore, there are observational indications that $w$
might evolve from $w>-1$ in the past to $w<-1$ at present, which
have stimulated many theoretical studies. For $w$ crossing $-1$, one
is encountered with the divergence problem for perturbations at
$w=-1$. For handling the parametrization of the EOS getting across
-1, firstly we introduce a small positive constant $\epsilon$ to
divide the full range of the allowed value of the EOS $w$ into three
parts: 1) $ w
> -1 + \epsilon$; 2) $-1 + \epsilon \geq w  \geq-1 - \epsilon$; and
 3) $w < -1 -\epsilon $.
Working in the conformal Newtonian gauge, the perturbations of DE
can be described by \bea
    \dot\delta&=&-(1+w)(\theta-3\dot{\Phi})
    -3\mathcal{H}(c_{s}^2-w)\delta~~, \label{dotdelta}\\
\dot\theta&=&-\mathcal{H}(1-3w)\theta-\frac{\dot{w}}{1+w}\theta
    +k^{2}(\frac{c_{s}^2\delta}{{1+w}}+ \Psi)~~ . \label{dottheta}
\eea

Neglecting the entropy perturbation, for the regions 1) and 3), the
EOS does not across $-1$ and the perturbation is well defined by
solving Eqs.(\ref{dotdelta},\ref{dottheta}). For the case 2), the
perturbation of energy density $\delta$ and divergence of velocity,
$\theta$, and the derivatives of $\delta$ and $\theta$ are finite
and continuous for the realistic quintom DE models. However for the
perturbations of the parameterizations, there is clearly a
divergence. In our study for such a regime, we match the
perturbations in region 2) to the regions 1) and 3) at the boundary
and set
\begin{equation}\label{dotx}
  \dot{\delta}=0 ~~,~~\dot{\theta}=0 .
\end{equation}
In our numerical calculations we limit the range to be $|\Delta w =
\epsilon |<10^{-5}$ and find our method to be a very good
approximation to the multi-field quintom. More detailed treatments
can be found in Ref.\cite{Zhao:2005vj}.

The publicly available Markov Chain Monte Carlo (MCMC) package
CosmoMC\cite{CosmoMC} is employed in our global fitting, and
modifications have been made to include DE perturbations, and to
suit the DE models which we study. We assume purely adiabatic
initial conditions and work in the flat universe with
$\Omega_{total}=1$. Our most general parameter space is:
\begin{equation}
\label{parameter} {\bf P} \equiv (\omega_{b}, \omega_{c},
\Theta_{s}, \tau,  w_{0}, w_{a}, n_{s}, \ln(10^{10}A_{s}))
\end{equation}
where $\omega_{b}\equiv\Omega_{b}h^{2}$ and
$\omega_{c}\equiv\Omega_{c}h^{2}$ are the baryon and cold dark
matter densities relative to the critical density, $\Theta_{s}$ is
the ratio (multiplied by 100) of the sound horizon at decoupling to
the angular diameter distance to the last scattering surface, $\tau$
is the optical depth due to re-ionization, $w_0$ and $w_a$ is the
parameters of the EOS of DE, $A_{s}$ and $n_{s}$ characterize the
power spectrum of primordial scalar perturbations. For the $\Lambda
CDM$, $w_0=-1, w_a=0$,

We vary the above parameters and fit to the observational data with
the MCMC method. For the pivot of the primordial spectrum we set
$k_{s0}=0.05$Mpc$^{-1}$. The following weak priors are taken: $\tau<
0.8$, $0.5<n_s<1.5$, $-3<w_0<3$, and $-5<w_a<5$. We impose a tophat
prior on the cosmic age as 10 Gyr $< t_0 <$ 20 Gyr. Furthermore, we
make use of the HST measurement of the Hubble parameter $H_0 = 100h
\quad {km\hspace{1mm} s}^{-1} {Mpc}^{-1}$
 by multiplying the likelihood by a Gaussian
likelihood function centered around $h=0.72$ with a standard
deviation $\sigma = 0.08$\cite{Hubble}. We also adopt a Gaussian
prior on the baryonic density $\Omega_{b}h^{2}=0.022\pm0.002$ (1
$\sigma$) from Big Bang Nucleosynthesis\cite{BBN}.

In our calculations, we take the total likelihood to be the products
of the separate likelihoods (${\bf \cal{L}}_i$) of CMB, LSS, SNIa
and GRB. Defining $\chi_{L,i}^2 \equiv -2 \log {\bf \cal{L}}_i$, we
then have \be\label{chi2} \chi^2_{L,total} = \chi^2_{L,CMB} +
\chi^2_{L,LSS} + \chi^2_{L,SNIa} +\chi^2_{L,GRBs}~ . \ee If the
likelihood function is Gaussian, $\chi^2_{L}$ coincides with the
usual definition of $\chi^2$ up to an additive constant
corresponding to the logarithm of the normalization factor of ${\cal
L}$. For CMB, we include the three-year WMAP (WMAP3) data and
compute the likelihood with the routine supplied by the WMAP team
\cite{Spergel:2006hy}. For LSS, we use the 3D power spectrum of
galaxies from SDSS \cite{Tegmark:2003uf} and from
2dFGRS\cite{Cole:2005sx}. To minimize the nonlinear effects, we
restrict ourselves only to the first 14 bins when using the SDSS
results\cite{sdssfit}, the range of k is $0.01578 < k/h <
0.10037$.  For SNe Ia, 
we mainly present the results with the recently released ``gold''
set of 182 supernovae published by Riess $et$ $al.$ in
Ref.\cite{Riess:2006}. For the GRB data, we have considered the
published sample by Schaefer\cite{shafer2006}, which includes 69 GRB
events. Upon using these distance modulus, we notice the circulation
problem associated with GRBs for cosmological probing. Due to the
lack of the local calibration, usually the correlations depend on
the cosmological parameters that we attempt to constrain. However,
there are results that show the relation does not change
dramatically in a wide range of cosmological
parameters\cite{firmani_circular}. In this paper we do not adopt the
correction for the circulation problem, and we take the distance
modulus published by Schaefer.  In the calculation of the likelihood
from SNe Ia and GRBs data, we marginalize over the nuisance
parameter\cite{DiPietro:2002cz, goliath2001}.

For each regular calculation, we run 8 independent chains comprising
of $150,000-300,000$ chain elements, and spend thousands of CPU
hours on a supercomputer. The average acceptance rate is about
$40\%$. We test the convergence of the chains by Gelman and Rubin
criteria\cite{R-1} and find that $R-1$ is on the order of $0.01$,
which is much more conservative than the recommended value
$R-1<0.1$.

\section{Results and Discussions}

\begin{table*}\label{table1}
Table I. Mean $1\sigma$ constrains on the EOS of DE and $\Omega_m$.
The left columns are obtained with
``WMAP3+SDSS+2dFGRS+SN${}_{gold}$+GRBs" combinations and the right
columns are correspondingly from without GRBs. 
\begin{center}
\begin{tabular}{c|ccc|ccc}

\hline
\hline

&\multicolumn{3}{c|}{WMAP3+LSS+SN${}_{gold}$+GRBs} &\multicolumn{3}{c}{WMAP3+LSS+SN${}_{gold}$}\\

\hline
 &$\Lambda$CDM&constant w& dynamical w(a)&$\Lambda$CDM&constant w& dynamical w(a)\\

\hline

$w_0$  & $-1$&$-0.853^{+0.077}_{-0.076}$&$-1.005^{+0.153}_{-0.151}$& $-1$&$-0.863^{+0.077}_{-0.076}$&$-1.001^{+0.166}_{-0.162}$  \\

$w_a$  & 0& 0&$0.533^{+0.454}_{-0.474}$& 0& 0&$0.443^{+0.527}_{-0.550}$\\

 $\Omega_m$
&$0.290^{+0.020}_{-0.020}$&$0.285^{+0.021}_{-0.020}$&$0.292^{+0.021}_{-0.021}$&$0.286^{+0.020}_{-0.021}$
&$0.283^{+0.020}_{-0.021}$&$0.288^{+0.021}_{-0.021}$\\



\hline \hline
\end{tabular}
\end{center}

\end{table*}

In this section, we present our global fitting results. We summarize
the $1\sigma$ constrains on the corresponding parameters in Table I.
We focus on the EOS of DE and $\Omega_m$. In order to show
explicitly the effects of GRBs, we compare the results between the
two cases with or without GRBs. From the Table, we can see that the
best fit values and the errors change when the GRBs data are
included, especially on the parameters of the EOS of DE. For
$\Lambda CDM$ or constant $w$ DE models, the best fit values and the
errors change little when considering the GRBs data, however, for
the dynamical DE models, the effects from GRBs can be obviously
seen. The best fit value of $(w_0, w_a)$ is $(-1.001, 0.443)$ for
the combined SN Ia $+$ WMAP3 $+$ SDSS $+$ 2dFGRS. When we adopt the
GRBs data, the best fit value of $(w_0, w_a)$ changed to $(-1.005,
0.533)$, and the error bars shrink considerably, especially on $w_a$
which delineate the evolution of DE. One can find that the $2\sigma$
error bar of $w_a$ changed from $0.443^{+0.747}_{-1.502}$ to
$0.553^{+0.663}_{-1.318}$, which is tightened about $15\%$. This can
be seen in Fig 1 graphically.

\begin{figure}[htbp]
\begin{center}
\includegraphics[scale=0.4]{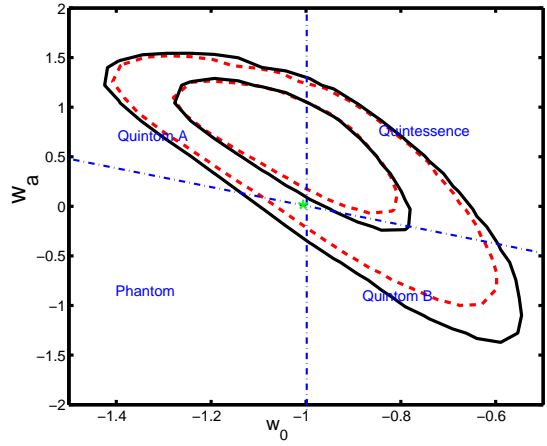}
\vskip-1.3cm \vspace{10mm}\caption{$2\sigma$ constrains in the $w_0$
and $w_a$ plane, the dotted red line is given by using WMAP3+182
"Gold" SNIa+LSS+GRBs with our 8-parameter parametrization discussed
in the text. Solid black curves come from without GRBs. For both
cases we considered perturbed dark energy. \label{w0-w1}}
\end{center}
\end{figure}

In Figure \ref{w0-w1}, we delineate the two dimensional posterior
constraint on $w_{0}-w_{a}$. The solid lines and the dashed lines
stand for $1\sigma$ and $2\sigma$ constraints respectively. We
divide the parameter space into four regions representing different
dark energy models by two lines: $w_0=-1$ and $w_0+w_a=-1$. For
models located in the upper left region labeled as Quintom A, the
equation of state of dark energy crosses $-1$ from upside down,
i.e., $w$ is greater than $-1$ in the past and becomes less than
$-1$ at present. The evolution of $w$ for models of Quintom B has an
opposite direction. From the plot, it is noted that Quintom A models
are mildly favored by current observational data. The $\Lambda CDM$
model continues to be a consistent one at $2\sigma$ level with and
without GRBs included.

\begin{figure}[htbp]
\begin{center}
\includegraphics[scale=0.8]{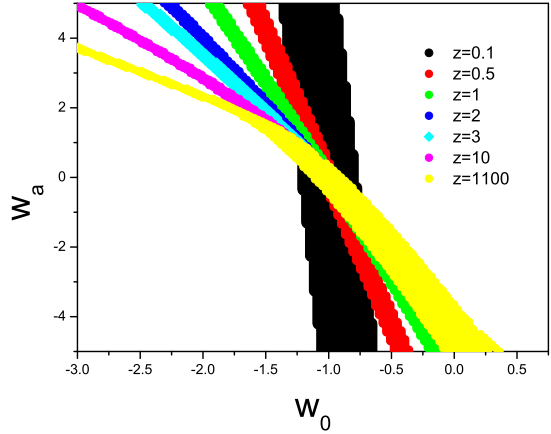}
\vskip-1.3cm \vspace{10mm}\caption{The different color region simble
$\pm 1\%$ variation around lines of constant $d_L$ at redshift $0.1$
(black), $0.5$(red),$1$(green),
$2$(blue),$3$(cyan),$10$(magenta),$1100$(yellow), taking $\Lambda$
CDM model as fiducial model. This plot delineate the degeneracy
between the parameters $w_0$ and $w_a$ at different redshift z.
\label{degeneracy}}
\end{center}
\end{figure}

The parameter $w_a$ represents the evolution of DE. It is known that
the CMB data alone cannot constrain well the dynamics of DE. LSS
data are valuable in DE studies mostly because they provide a tight
limit on $\Omega_m$, which in turn helps to constrain the properties
of DE due to the degeneracy between $\Omega_m$ and the EOS of DE in
cosmological observables. Currently the measurements of the
luminosity-distance from SNe Ia at various redshifts give rise to
the most direct constraints on the dynamics of DE. In a flat
universe with the EOS of DE given by Eq. (1), the luminosity
distance can be written as \be d_L=c(1+z)\int^{z}_{0}\frac{d
z^{\prime}}{H_0[\Omega_m
(1+z^{\prime})^3+(1-\Omega_m)(1+z^{\prime})^{3(1+w_0+w_a)}
\exp{[-3w_a\frac{z^{\prime}}{1+z^{\prime}}]}]^{\frac{1}{2}}}.\ee
From this equation, one can see that there are degeneracies between
the background parameters $w_0$, $w_a$, $\Omega_m$ and $H_0$. In
figure \ref{degeneracy} we present the degeneracy by exploring the
information of luminorsity distance at different
redshifts in the $w_0-w_a$ parameter plane. 
Different colored bands describe the parameter space of $w_0, w_a$
where the variation of $d_L$ is in between $\pm 1 \%$ for $z=0.1$
(black), $0.5$(red),$1$(green), $2$(blue),$3$(cyan),$10$(magenta)
and $1100$(yellow). One can find that the degeneracy between $w_0$
and $w_a$ varies with the redshift, which in turn implicit that
combining the information of $d_L$ at different redshifts can help
break such a degeneracy. Therefore, in order to constrain the
cosmological parameters well, one needs distance determinations for
a wide range of redshifts. For current SNe Ia data, the redshift
range is limited with the highest observed redshift $\sim 1.7$. On
the other hand, for the GRB sample used in our analysis, the
redshift extends to as high as $\sim 6.3$ with $39$ data points
having $z>1.5$, thus, the complementarity of GRBs to SNe Ia is
highly expected. This plot is the ideal case for showing the
degeneracy between the parameters $w_0$ and $w_a$ for different
redshift, because we fix the other parameters except $w_0$ and
$w_a$. In fact, if we vary the other parameters like $\Omega_m$ or
$H_0$, the changing of the degeneracy with redshift will also exist
but will not be so obvious as figure \ref{degeneracy}, the different
colored bands will be much broader than the ones in figure
\ref{degeneracy} respectively. Figure \ref{w0-w1} is more general
results which is obtained from the global fitting where we have all
the parameters in equation (\ref{parameter}) varying.

\begin{figure}[htbp]
\begin{center}
\includegraphics[scale=0.4]{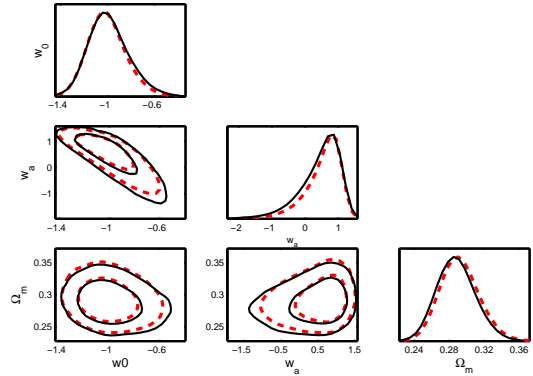}
\vskip-1.3cm \vspace{10mm}\caption{1-d posterior constraints and 2-d
joint $68\%$ and $95\%$ confidence regions for the parameters $w_0$,
$w_a$ and $\Omega_m$ obtained via MCMC methods. The dotted red line
is given by using WMAP3+182 "Gold" SNIa+LSS+GRBs while the solid
black curves come from without GRBs. For both cases we considered
perturbed dark energy. \label{triplot}}
\end{center}
\end{figure}

To demonstrate the degeneracies between different parameters, we
show in Figure \ref{triplot} the two-dimensional correlation
contours of $w_0$, $w_a$ and $\Omega_m$ from our global fitting. It
is seen that, with the GRBs data, the constraints on the parameters
are tightened to a certain degree, especially for $w_0$ and $w_a$.
Our results show that high-redshift cosmological probes can play
roles in the study of the dynamics of dark energy, and thus it is of
importance to explore high-redshift cosmological probes.

GRBs can bridge up the gap between the relatively nearby SN Ia and
the much earlier CMB. Comparing with SN Ia, GRBs have their own
advantages. They are very powerful and thus can be detected out to
very high redshifts. Furthermore, the gamma-ray photons suffer from
no dust extinction when they propagate to us. Therefore they deserve
detailed studies. Although the cosmological applications of GRBs are
currently limited by the relatively small number of available data
and the quality of the data, it is considerably important to
investigate their potentials and related problems. With the
accumulation of observational data and the advances of theoretical
understandings of GRB physics, the goal of using GRBs as
cosmological candles could be better achieved.

\section{Summary}
In this paper, we have made the first global fitting with the
combined GRB, WMAP3, SDSS, 2dFGRS and SN Ia data. We concentrate on
dynamical DE models, and explore the complementarity of GRBs to
other cosmological probes. DE perturbations are treated carefully.
From the global fitting results given by MCMC, we find that
high-redshift GRBs may have effects in constraining the EOS of DE,
especially for the dynamical DE models. Including GRBs data can
shrink the contours and modify the best fit values of DE parameters.
The $\Lambda$CDM model is consistent with the current data within
$2\sigma$ confidence level, with Quintom A-like models mildly
favored.

GRBs are the most powerful astrophysical events in the universe,
which hold great potential to reveal the high redshift universe.
Even though, the GRB data are not as good as SN Ia currently, our
results indicate the possible potentials of GRBs in dark energy
studies.

It is worth mentioning that our global fitting method is different
from using the CMB shift parameter, the linear growth factor or the
BAO parameter measured from the LSS. Although these parameters carry
important information of dark energy, their specific values are
usually extracted from observational data under certain conditions.
Overlooking these conditions leads to inappropriate use of the
values of these parameters, which in turn could result biased
constraints on cosmological parameters. Our MCMC analysis are
performed on observational data directly and therefore avoid such
problems. We also take into account dark energy perturbations, whose
effects are not included in the shift parameter and the BAO
parameter.

The analyse presented in this paper mainly aim at emphasizing the
importance of high-redshift cosmological probes, which are not
limited to GRBs. Our results demonstrate their contributions to dark
energy studies clearly, especially on the dynamics of the dark
energy component.

\acknowledgments

Our MCMC chains were finished in the Sunway system of the Shanghai
Supercomputer Center(SSC). Hong Li thanks the Center de Physique des
Particles de Marseille for the kind hospitality during part of the
preparation of this work and acknowledge the funding support from
the china postdoctoral science foundation. This work is supported in
part by National Science Foundation of China under Grant Nos.
90303004, 19925523, 10243006, 10373001, 10233010, 10221001 and
10533010, and by Ministry of Science and Technology of China under
Grant No. NKBRSF G19990754 and TG1999075401, by the Key Grant
Project of Chinese Ministry of Education (No. 305001). We thank
Junqing Xia, Gong Bo Zhao, Bo Feng, Jingsong Deng, Charling Tao and
F. Virgili for helpful discussions.

\end{document}